\RequirePackage{lineno}
\documentclass[prb,twocolumn,aps,amsmath,amssymb,floatfix]{revtex4}
\usepackage{graphicx}% Include figure files                                                                                                           
\usepackage{bm}% bold math                                                                                                                
\usepackage{ulem}
\usepackage{color,graphicx,pstricks}
\begin{document}

\title{Antiferromagnetism beyond classical percolation threshold in the site-diluted
  half-filled one-band Hubbard model in three dimensions}

\author{Sourav Chakraborty$^{1}$, Anamitra Mukherjee$^{2,}$\footnote{anamitra@niser.ac.in}, 
Kalpataru Pradhan$^{1,}$\footnote{kalpataru.pradhan@saha.ac.in}}
\affiliation{$^{1}$Theory Division, Saha Institute of
Nuclear Physics, HBNI, Kolkata-700064, India \\
$^{2}$School of Physical Sciences, National Institute of Science Education and Research, 
HBNI, Jatni 752050, India}

\date{\today}

\begin{abstract}
We investigate the impact of site dilution by setting the on-site repulsion
strength ($U$) to zero at a fraction of sites in the half-filled Hubbard model
on a simple cubic lattice. We employ a semi-classical Monte-Carlo approach first
to recover the zero dilution (undiluted $x=1$) properties, including $U$
dependence of insulator to metal crossover temperature scale $T^*$ and long-range
staggered antiferromagnetic ordering temperature ($T_N$). For the non-perturbative
regime of $U \sim$ bandwidth, we find a rapid suppression of $T^*$ with reducing
$x$ from 1 to 0.7. However, $T_N$ remains unchanged in this dilution range, showing
a weakening of the insulating state but not of the magnetic order. At $x \leq 0.7$,
$T^*$ and $T_N$ coincide and are suppressed together with further increase in
site-dilution. Finally, the system loses the magnetic order and the insulating state
for $x=0.15$, significantly below the classical percolation threshold $x_p^{sc} (\sim 0.31$). 
We show that the induced moments on $U=0$ sites drive the magnetic order below the
classical percolation limit by studying local moment systematics and finite-size
analysis of magnetic order. At the end, we show that either increasing $U$ to large
values or raising temperature beyond a $U$ dependent critical value,  suppresses the
induced local moments of the $U=0$ sites and recovers the classical percolation
threshold.

\end{abstract}

\maketitle 

\section{Introduction}

The discovery of high-temperature superconductivity in doped copper oxides generated
an enormous amount of interest in quantum antiferromagnets~\cite{anderson, sachdev, keimer}.
The emergence and collapse of long-range antiferromagnetic (AF) order, which provides
us a unique way to explore many exotic magnetic phases, is one of the most essential
and well-explored topics in condensed matter physics. The AF ordering in cuprate,
iron-pnictide, and iron-chalcogenides gets suppressed by doping non-magnetic
impurities~\cite{liu1,liu2, norman}. Spin-wave theory for low concentration of
impurities with the impurities treated as static vacancies~\cite{kircan,chernyshev}
can usually model such behavior. On the other hand, frustration arising from in-plane
couplings in clean systems can also disrupt long-range magnetic order (LRO) and are
routinely explored within the $J_1-J_2$ Heisenberg spin models~\cite{sato,wang,schmidt}.
Disorder-induced suppression of long-range magnetic order~\cite{szczech,kunwar}
have been typically studied in the strong-coupling limit. Quantum Monte Carlo
simulations~\cite{sandvik} in the large correlation strength limit, agree with the
AF order vanishing at the classical percolation threshold as in the experiments.

Cuprates like La$_{2}$Cu$_{x}$Mg$_{1-x}$O$_{4}$~\cite{mccarron,cheong,wan,vajk}
have inspired some site-diluted Hubbard model studies in two and quasi-two
dimensions~\cite{ulmke,medhi,delannoy}. The experimental motivation was to
investigate superconductivity in the parent material (La$_{2}$CuO$_{4}$) by
doping with non-magnetic Zn or Mg that suppresses long-range antiferromagnetic
order. According to the current understanding this quasi-two dimensional
material [La$_{2}$Cu$_{x}$(Mg/Zn)$_{1-x}$O$_{4}$] shows complete suppression of
long-range AF order~\cite{vajk} for $x_p^{2D} \sim 0.59$, the classical percolation
threshold~\cite{christensen,sandvik}. In the strong interaction limits for such
materials where 3$d$ transition metal elements are
involved~\cite{sandvik,kato, chernyshev1, carretta}, the long-range magnetic
order vanishes at $x_p$, the critical classical percolation threshold in
the relevant dimensions.

However, investigations of site dilution for the Hubbard model where $U$ is
comparable to bandwidth (BW) are relevant both for materials and is of
theoretical interest. In particular, since the correlation-induced suppression
of double occupation is not too severe, sites with $U=0$ in vicinity of
$U\neq0$ sites can get effected by virtual charge fluctuations leading to
induced moments on the uncorrelated sites. Thus, whether site dilution will
suppress the long-range antiferromagnetic order of the undiluted system is
unclear. A recent study of the diluted Hubbard model on Lieb lattice shows
that magnetic order is very robust for dilution much lower than the classical
percolation threshold~\cite{lima}.

In this paper, we analyze the effect of site dilution in a Hubbard model
at half-filling using a semi-classical Monte-Carlo scheme (s-MC)~\cite{mukherjee,ptca-ref}.
The method reduces to an unrestricted Hartree-Fock method at very low
temperatures but becomes progressively accurate with temperature increase and,
in particular, compares well with Determinant Quantum Monte Carlo (DQMC) over
a wide temperature range. We consider the half-filled Hubbard model as defined
below in three dimensions. We first produce several benchmarks for the undiluted
case, including the AF magnetic order. We then show that switching off interaction
potential on a fraction of sites weakens and eventually destroys the magnetic
order. However, remarkably, for correlation strength, where the bandwidth (BW)
and interaction strength ($U$) are comparable, we show that the AF order survives
to dilutions much below the classical percolation threshold. We investigate this
phenomenon by tracking the local moment dependence with temperature. We show that
$U\neq0$ sites induce significant suppression of double occupation on the $U=0$
sites stabilizing local moments on the uncorrelated sites. In addition our
calculations reveal that the density of states carries the signature of this effect
and manifests as a four-lobe Mott insulator. At a critical dilution below the
classical threshold, we show that the collapse of local moments at the $U=0$
sites signals the onset of a metallic state. We find that the ensuing metal has
a pseudo-gapped density of states at low temperatures. We characterize the
percolative metallic state and it's temperature dependence. Finally, we demonstrate 
that the vanishing of the AF order at the classical percolation threshold occurs
for $U$ much larger than the BW, in agreement with earlier literature. We also
find that the same can happen at increased temperatures where thermal fluctuations
destroy the local moments on the $U=0$ sites. Thus we present a complete
phenomenology within our semi-classical approach of site dilution effects,
correlation strength, and temperature.

\section{Model \& Method}

We consider the following particle-hole symmetric from of the one band Hubbard Hamiltonian:

{\small
\begin{eqnarray}
  H=-t\sum_{<i,j>,\sigma}(c^{\dagger}_{i\sigma}c_{j\sigma} + h.c.)~~~~~~~~~~~~\nonumber\\+ U\sum_{i} \Big(n_{i\uparrow}-\frac{1}{2}\Big) \Big(n_{i\downarrow}-\frac{1}{2}\Big)
  -\mu \sum_{i} n_i 
\end{eqnarray}

\noindent
where $c_{i\sigma}$ ($c^{\rm \dagger}_{i\sigma}$) are the fermion annihilation
(creation) operators at site $i$ with spin $\sigma$. $t$ is the nearest neighbor
hopping parameter and $U$ ($>$ 0) denotes the on-site repulsive Hubbard interaction. 
$\mu$ is the chemical potential.

To employ the s-MC approach we decompose the on-site interaction term by introducing
standard Hubbard-Stratonovich (HS) auxiliary fields (a vector field $m_i$ that couples
to spin degrees of freedom while a scalar field $\phi_i$ couples to charge degree of
freedom) at each site $i$. We treat auxiliary fields as {\it classical fields} by
dropping the time dependence explicitly. We treat $\phi_i$ at the saddle point level
$i\phi_{i} = \frac{U}{2} \textless n_{i} \textgreater $, but retain the
thermal fluctuations for $m_i$. These thermal fluctuations are necessary 
to capture many of the well established features which will be discussed later.
The following effective spin-fermion Hamiltonian is derived using above
approximations (see supplementary materials for the details and justification
of the approximation):

\begin{eqnarray}
H_{eff}= -t\sum_{<i,j>,\sigma} (c^{\dagger}_{i\sigma}c_{j\sigma} + h.c.)\nonumber 
+ U/2 \sum_{i} (\textless n_i \textgreater n_i - \textbf{m}_i . \sigma_{i})  \nonumber \\
  + (U/4)\sum_{i}(\textbf{m}_{i}^2 - {\textless n_i \textgreater}^2) -\frac{U}{2} \sum_{i} n_{i}  -\mu \sum_{i} n_i \nonumber
\end{eqnarray}

The chemical potential is varied to maintain the system at half filling.
We solve $H_{eff}$ by using exact diagonalization based Monte-Carlo method.
We diagonalize the Hamiltonian for a fixed set of $\{\textbf{m}_{i}\}$
and $\{\textless n_i \textgreater\}$ configuration. We update the $\{\textbf{m}_{i}\}$
at each site based on usual Metropolis scheme at a fixed temperature.
The $\textless n_i \textgreater$ fields are self consistently updated at every
10$^{th}$ Monte-Carlo step where the $\textbf{m}_{i}$ fields are held fixed. The
goal of the process is to generate equilibrium configuration of the $\textbf{m}_{i}$ and
the $\textless n_i \textgreater$ fields. Expectation values of observables are obtained
by appropriately using the eigenvectors and eigenvalues resulting from diagonalizing
the Hamiltonian in each of the equilibrium configuration. These individual expectation
values from equilibrium configurations are further averaged over results from 100
such configurations at a fixed temperature. All observables are calculated at a given
temperature by averaging over the values obtained from individual configurations. We
note that we calculate observables from every tenth equilibrium configuration to
avoid spurious self correlations. Temperature is lowered in small steps to allow
for equilibration. To avoid size limitation we employ above mentioned Monte-Carlo
technique within a traveling cluster approximation~\cite{tca-ref,pradhan,chakraborty} 
to handle system size $N$ = $L^3$ = 10$^3$. 

For $0<x<1$, we have $Nx$ fraction of sites with finite $U$ and $N(1-x)$ sites with
$U=0$, with $N$ being the total number of sites. For $U=8$, on-site interactions
($U_i$) at each sites are chosen using the distribution
$P(U_i)=(1-x)\delta(U_i)+x\delta(U_i-8)$. We
introduce the HS auxiliary fields only on the $U\neq0$ sites. The induced moments
on the $U=0$ sites are calculated by computing quantum local moments as we discuss
later. Finally, all energy scales ($U$, temperature, BW etc.) are measured in units 
of the hopping parameter $t$.

\section{Phase Diagram}

First we discuss briefly about the $U-T$ phase diagram, seen in the inset of 
Fig.~\ref{fig_pd}, obtained for $x$=1 (i.e. the undiluted case, without any
$U$=0 sites). We first find that for all $U$, we have a staggered AF insulating
ground state (AF-I). The staggered AF transition temperature $T_N$ defines
the finite temperature boundary of the AF and the PM phase. The antiferromagnetic
transition temperature $T_N$ increases with $U$ up to $U$=8 and decreases
thereafter. For larger $U$ s-MC captures $\sim t^{2}/U$ scaling of $T_N$. In
the inset we also find that for large $U$, above $T_N$, there is an insulating
region of preformed local moments with no long range magnetic order (PM-I).
This phase crosses over to a paramagnetic metal (PM-M)
above the dashed line. We discuss below how these phases are determined for
different values of $x$. For the $x=1$ case, the non-monotonic $U$ dependence
of $T_N$ and the preformed local moment regime at finite temperature are results
beyond simple finite temperature Hartree-Fock mean field theory. Details and
comparison with DQMC are presented in earlier s-MC literature~\cite{mukherjee}.
s-MC has also been used to study the physics of Anderson-Hubbard model~\cite{patel}
and frustrated Hubbard model~\cite{gour, tiwari}.

% -----------------------------------------------------                                         
\begin{figure}[t]
\centerline{
\includegraphics[width=9.0cm,height=7.7cm,angle=0,clip=true]{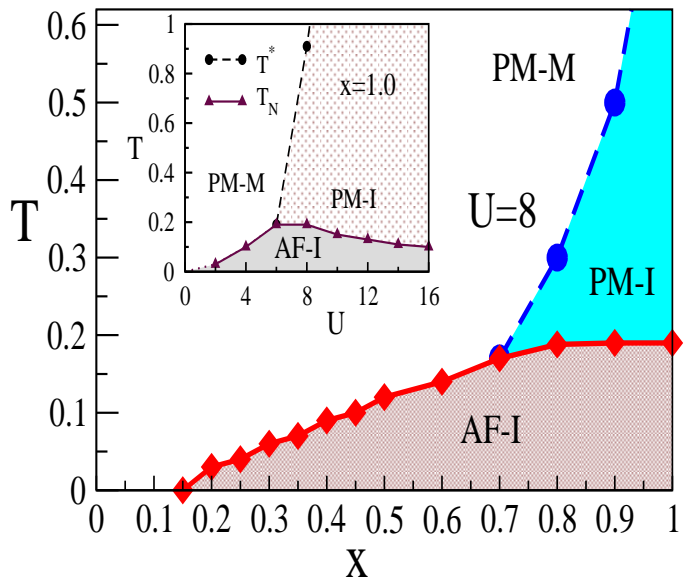}}
\vspace{.2cm}
\caption{\label{fig_pd} $x-T$ phase diagram for $U$=8. $x$ is the concentration
  of correlated sites ($U$=8) and rest of the sites (with concentration $1-x$)
  have $U=0$. PM-I phase intervenes between the PM-M and AF-I phase for $x$ $>$ 0.7.
  For $x \leq$ 0.7 the PM-I phase vanishes and the metal insulator transition coincides
  with the onset of the AF order. For $x \leq$ 0.15 the AF order completely collapses at low
  temperatures. The inset shows the $U-T$ phase diagram. For details please
  see the text. 
}
\end{figure}
% -----------------------------------------------------

As mentioned in the introduction, our main motivation is to examine if the AF order
can survive below the classical percolation threshold. For this we initially confine
to $U=8$ where $T_N$ is optimum and study the effect of site-dilution. This value
of $U$ is away from the two perturbative limits of $U/BW<<1$ and $U/BW>>1$. We will
discuss the systematics of  varying $U$ at a later stage. 
The main panel in Fig.~\ref{fig_pd}, shows the $x-T$ phase diagram for $U=8$. We
see that the $T_N$ (diamonds) and crossover scale (dashed line) both decrease as $x$
is reduced. Within numerical accuracy, the PM-I phase exists for $x>0.7$. The important
observation is that the $T_N$ survives up to $x=0.15$, much smaller than the classical
three dimensional percolation threshold ($x_p^{sc} \sim 0.31$). We will show below that
this conclusion is robust to changes in the system size.

% -----------------------------------------------------                                            
\begin{figure}[t]
\centerline{
\includegraphics[width=9.0cm,height=7.7cm,angle=0,clip=true]{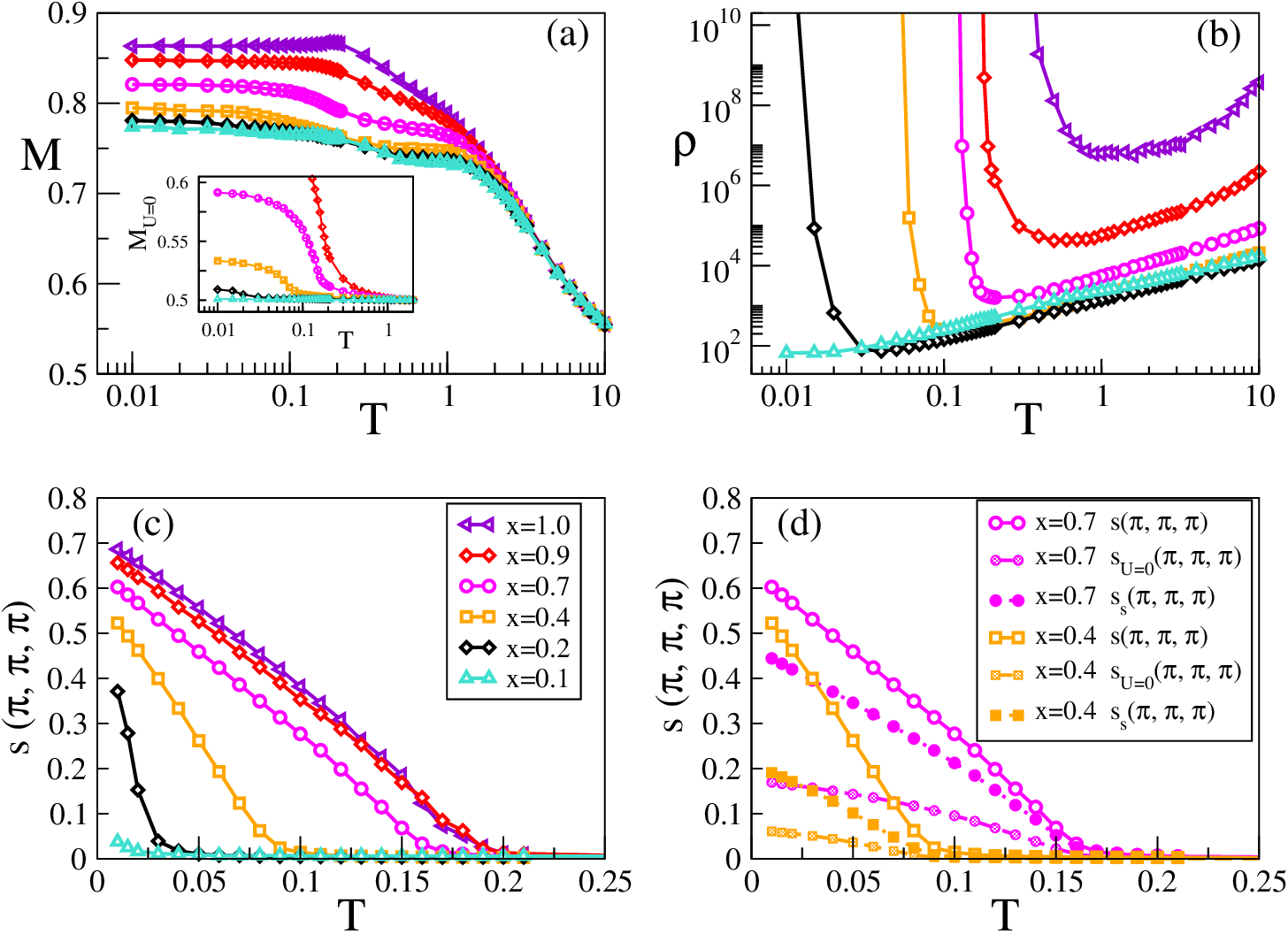}}
\vspace{.2cm}
\caption{\label{fig_af}
Physical quantities for different $x$ values. All calculations are
done for $U=8$ unless otherwise specified.
(a) Average local moments $M$ vs temperature measured using only $U=8$ sites.
Inset shows the induced moments with temperature at U=0 sites. Legends are same
in (a)-(c). (b) Resistivity vs temperature curves for different $x$ values.
(c) s($\pi, \pi, \pi$) vs temperature (by taking $U=8$ sites only) for
different $x$. With decrease in $x$ the quantum structure factor decreases at low
temperature. $T_N$ also decreases and vanishes for $x$ = 0.1.
(d) $s_s$($\pi, \pi, \pi$), structure factor for taking both $U=0$ and $U=8$ sites,
vs temperature shows that the system as a whole turns antiferromagnetic
at the same temperature to that of $U=8$ sites. For comparison s$_{U=0}$($\pi, \pi, \pi$),
calculated only for $U=0$ sites for specific $x$ are also shown.
}
\end{figure}
% -----------------------------------------------------

\section{Magnetic and Transport Properties}

Next we discuss the magnetic and transport properties in details that
we used to construct the phase diagram for site-diluted Hubbard model
(Fig. \ref{fig_pd}). 

\textit{Metal-insulator transition \& magnetic order:} From the $x=1$ analysis
we know that local moments exist on all sites for $U=8$.  For $x=1$ the system
averaged quantum local moment is defined as $M= \langle (n_{\uparrow} - n_{\downarrow})^{2} \rangle = 
\langle n \rangle - 2 \langle n_{\uparrow} n_{\downarrow} \rangle$, where
the angular brackets imply quantum and thermal averaging at individual sites.
The large moment at low temperature is due to the increased suppression of
doublons (local double occupation). In the limit $U\rightarrow \infty$ and any
finite temperature we expect the double occupation to go to zero giving $M=1$.
In the other extreme limit of $U=0$ or $T\rightarrow\infty $ at any finite $U$,
the double occupation
$\langle n_{\uparrow} n_{\downarrow} \rangle\rightarrow \langle n_{\uparrow}\rangle\langle n_{\downarrow}\rangle=0.25$.
This gives the value of $M$ to be 0.5. In Fig.~\ref{fig_af} (a) main panel we
show the local moment $M$ vs temperature data for various $x$ values for the
$U\neq0$ sites. For $x=1$, this of course coincides with the system averaged
local moments, while for $x\neq 1$, the site averaging is done only over the
$U\neq 0$ sites.

To understand the actual temperature scale for moment formation that lies between
these two limits and its impact on transport properties we calculate the resistivity
as function of temperature in Fig.~\ref{fig_af}(b). Resistivity is calculated for
different $x$ values by calculating the {\it dc} limit of the optical conductivity
determined by the Kubo-Greenwood formula~\cite{mahan-book,cond-ref}. A metal to insulator
crossover (MIC) scale ($T^*$) is ascertained from the sign of $d\rho/dT$. For $x=1$,
and $U=8$, $T^*\sim 1$. For the $x<1$ case we see that this crossover scale reduces
rapidly. In order to understand this systematics, in (a) we plot the local moments
of the correlated ($U\neq 0$) and uncorrelated ($U= 0$) sites separately. For the
correlated sites we see that there is an overall reduction in the local moment
magnitudes, but there is no drastic local moment collapse to suggest a significantly
smaller MIC temperature as the resistivity data suggests. However, the remarkable
effect on deciding the scale for onset of metallicity with temperature increase
comes from the $U=0$ sites! In the inset of (a), we see that weak local moments
are induced on the uncorrelated sites. We observe that $T^*$ is controlled  by the
onset temperature of the local moments formation ($M$ becoming greater than 0.5) 
on the uncorrelated sites. For example for $x=0.9$, the local moments of the
correlated sites are similar in magnitude to the $x=1$ case, yet the resistivity data
shows an insulator to metal crossover occurs at  $T^*\sim 0.5$ as opposed to $\sim$1
for $x=1$ case. Interestingly the onset of magnetic moments on $U=0$ sites,
transition from PM-M to AF-I phase occurs at same temperature for $x\leq 0.7$.
This clear correlation needs the following clarification: whether the small moments
on the $U=0$ sites, for example small moments very close to the uncorrelated value
of 0.5, for $x < 0.4$, responsible for stabilizing the low temperature antiferromagnetic
insulating state? In particular, is there a long range order arising out of the sites
with two values of moments?

To answer this question, in (c) we plot the quantum antiferromagnetic correlations
s($\pi, \pi, \pi$) [$s(\bold{q})=\frac{1}{(Nx)^{2}} \sum_{mn} \langle \bold{s}_{m}.\bold{s}_{n}\rangle e^{i\bold{q}.(r_{m}-r_{n})}$
where $\bold{q}$ is the wave vector], where angular brackets have the same meaning as
mentioned above and the indices $\{m,n\}$ run over only the $U=8$ sites. The normalization
is defined accordingly. The reduction in the $T_N$ as well as the low $T$ saturation value
with decreasing $x$ is apparent. The weakening of antiferromagnetism due to site dilution 
is expected. However, we find an unexpected behavior when comparing the above with magnetic
structure factor computed only for the $U=0$ sites [denoted as s$_{U=0}$($\pi, \pi, \pi$)].
In (d) we show two such comparisons at the indicated values of $x$. The AF order that
results from the $U=0$ sites by itself generates long range staggered AF correlations
in three dimensions. Further, the AF order from the two set of sites (taking U=8 and U=0
sites separately) and AF correlations s$_{s}$($\pi, \pi, \pi$) obtained by taking all the
sites (both  U=8 and U=0 sites at the same time) vanish at the same temperature. This
shows cooperation of the $U\neq 0$ and $U=0$ matrix. In addition, for $x \leq 0.7$,
we find that the insulating state and the $T_N$ coincide, reminiscent of a Slater like
insulator. However, unlike the Slater insulator, this is clearly not arising out of a
nesting instability.  For $x$ $>$ 0.7, the data suggests that the insulting state can
survive without the magnetic order, but requires finite local moments at the
$U=0$ sites. This is the continuation of the $x=1$ PM-I phase for $x<1$.

% -----------------------------------------------------                                                        
\begin{figure}[t]
\centerline{
\includegraphics[width=9.0cm,height=7.7cm,angle=0,clip=true]{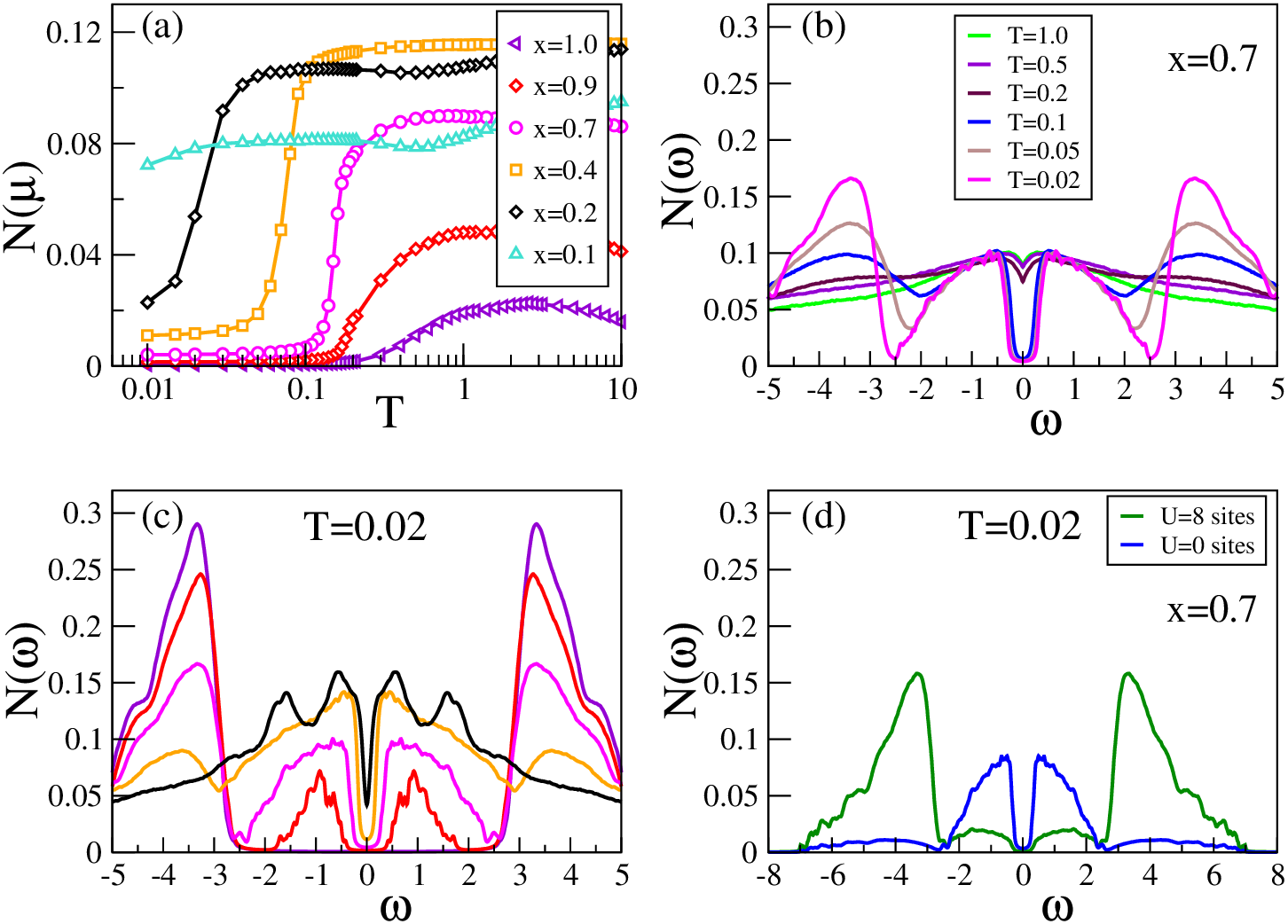}}
\vspace{.2cm}
\caption{\label{fig_mit}
(a) Temperature dependence of density of states at the Fermi energy N($\mu$)
shows that the Mott gap collapses at high temperature due to thermal fluctuation.
Thermal fluctuation persists up to low enough temperature for lower $x$ values.
N($\mu$) more or less remains constant with temperature for $x=0.1$.
(b) Density of states N($\omega$) with $\omega$ for different temperatures
at $x=0.7$. (c) Density of states N($\omega$) with $\omega$ for different $x$ 
at $T=0.02$. Legends are same in (a) and (c).
For $x=1$, the Mott gap at $T=0.02$ show Mott lobes around 
$\pm U/2$. A pair of secondary Mott lobe forms near $\omega=0$ for $x<1$.
(d) $U$ dependent density of states for $x=0.7$ shows that the secondary lobes are
mainly due to the $U=0$ sites.
}
\end{figure}
% -----------------------------------------------------

\textit{Density of states:} In  Fig.~\ref{fig_mit} (a) we show the temperature
evolution of the density of states at the chemical potential N($\omega=\mu$) for
different values of $x$. Density of states are obtained by implementing the Lorentzian
representation of the $\delta$ function in $N(\omega) =  \sum_{k} \delta(\omega - \omega_{k})$,
where $\omega_{k}$  are the eigenvalues of the fermionic sector and the summation
runs up to $2L^{3}$, i.e. the total number of eigenvalues of a $L^{3}$ system.
The expected gradual filling up of the charge gap in the Mott state seen for
$x=1$ as also seen in~\cite{dqmc} for smaller $x$ values. The gap filling however
becomes abrupt for $x\leq 0.7$. This is the same dilution below which we have a
direct transition from a PM-M to an AF-I. In addition, the density of states are
plotted explicitly for $x=0.7$, in Fig.~\ref{fig_mit}(b) at different temperatures.
While not explored herein detail, overall the data suggests a possible first 
order transition for $x\leq0.7$. Finally for $x=0.1$, we have a gap-less ground state. 

The density of states are compared for different $x$ values at low temperature
$T=0.02$, in Fig.~\ref{fig_mit}(c). We see that with reducing $x$, the upper and
lower Hubbard sub-bands around $\pm U/2$ evolves in to a four sub-band structure and
the gap around chemical potential $\mu$ ($\omega$=0) reduces. The gap eventually 
closes and we find a pseudo-gapped metal for $x=0.1$. 
To understand the origin of the new features in the density of states we show the
contribution of the DOS from $U=0$ and $U=8$ sites separately in panel (d) for
$x=0.7$ at the same low temperature. It shows that $U=0$ sites mostly contribute
to the formation of the low energy Mott lobes (around $\omega=\pm1$). This four
sub-band DOS within s-MC qualitatively agrees with DQMC~\cite{ulmke}.

% -----------------------------------------------------
\begin{figure}[t]
\centerline{
\includegraphics[width=9.0cm,height=7.7cm,angle=0,clip=true]{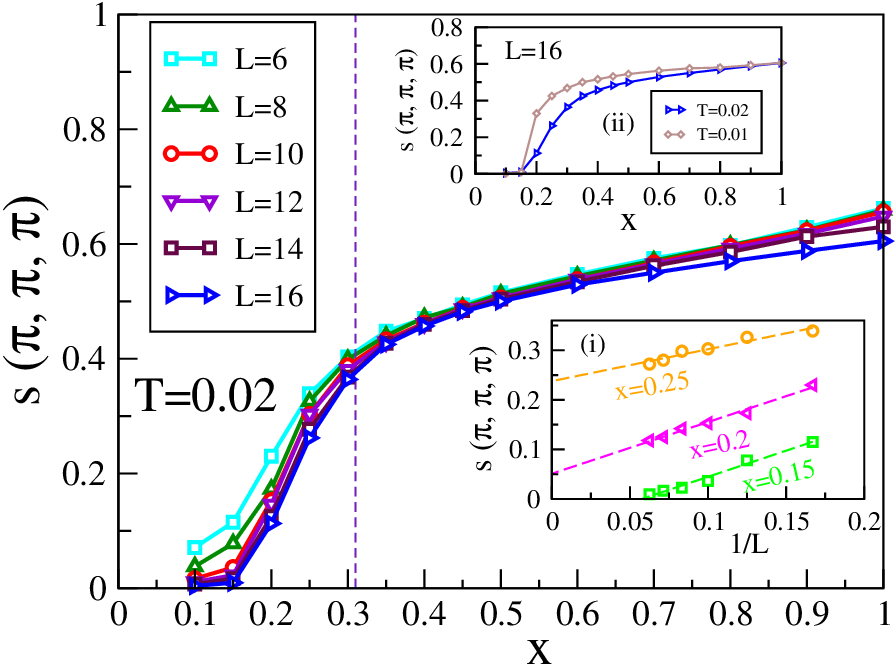}}
\vspace{.2cm}
\caption{\label{fig_pc}
Shows the drop of s($\pi, \pi, \pi$) with decrease in $x$ values for different
L values (N=L$^3$). For
$T=0.02$ long range AF order persists up to $x$=0.15. This AF order is due
to the induced moment formed at $U=0$ sites as explained the text. The
result is almost indistinguishable beyond system size $10^3$.
All calculations are done for $U=8$. Inset(i) shows the s($\pi, \pi, \pi$)
vs $1/L$ for $x=0.15, 0.2$ and $0.25$. Inset also shows linear fitting of
s($\pi, \pi, \pi$) with $1/L$ for $x$ = 0.15, 0.2 and 0.25. Inset(ii)
shows that the long range AF order also persists up to $x$=0.15 for $T=0.01$.
}
\end{figure}
% -----------------------------------------------------

% -----------------------------------------------------                                               
\begin{figure}[t]
\centerline{
\includegraphics[width=9.0cm,height=7.7cm,angle=0,clip=true]{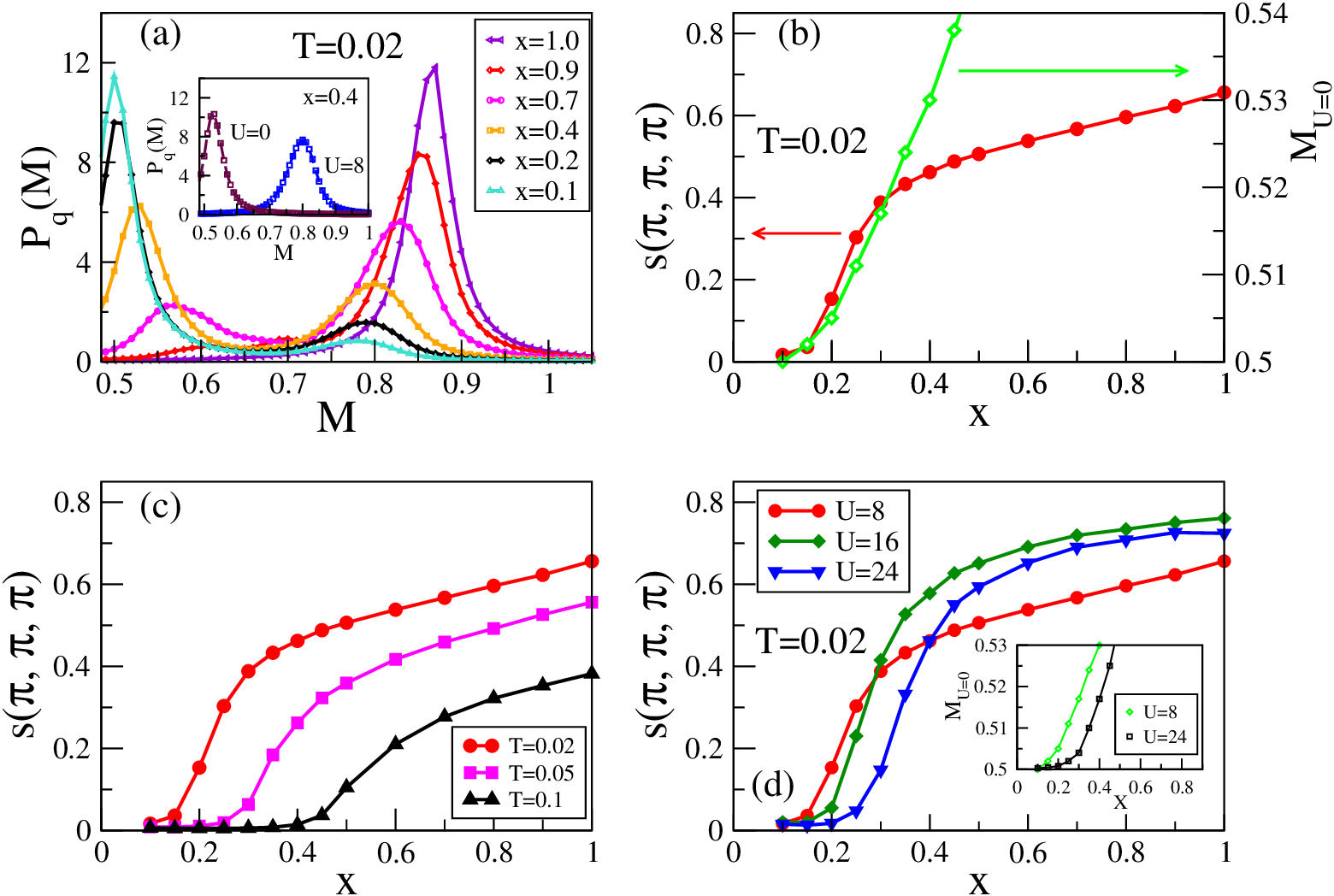}}
\vspace{.2cm}
\caption{\label{fig_mom}
(a) Distribution of local moments at $U=8$ and $U=0$ sites for
different $x$ values; the peak around $M=0.8$ is for the finite $U$ sites
whereas peak at lower $M$ is for the induced moments ($M_{U=0}$) at $U=0$
sites. Inset shows the distribution for $U=0$ and $U=8$ sites separately
at $x=0.4$.
(b) s($\pi, \pi, \pi$) and induced moment ($M_{U=0}$) at $U=0$ sites
with $x$ for $T=0.02$ shows one to one correspondence between the onset
of antiferromagnetic correlations and formation in induced magnetic moments
at $U=0$ sites. The induced moments mediates the antiferromagnetic correlation
below classical percolation threshold limit. 
(c) s($\pi, \pi, \pi$) vs $x$ for different temperatures show that 
the percolation threshold increases with temperature.
(d) s($\pi, \pi, \pi$) vs $x$ for different $U$ values at $T=0.02$.
The percolation threshold increases with $U$ values due to suppression
of charge fluctuations at large $U$. The inset shows comparison of
the induced moment ($M_{U=0}$) at $U=0$ sites with $x$ between $U=8$ and
$U=24$.
}
\end{figure}
% -----------------------------------------------------

\section{Percolation Threshold}

To start with, we first demonstrate the stability of the  
critical percolation threshold obtained from finite size lattices. Since 
the AF order parameter is the central indicator used by us, in Fig.~\ref{fig_pc} 
we show the low temperature value of $s$($\pi, \pi, \pi$) for different system 
sizes as a function of $x$. The system size is defined as $L^3$. Beyond $10^3$
the results for low $x$ are barely distinguishable from each other. We see from
the data that the order parameter rapidly converges with system size, giving
the limiting value of $x_p=0.15$. The inset(i) shows  $s$($\pi, \pi, \pi$) plotted
as a function of inverse system size ($1/L$) for three values of $x$. We find
the for $x>0.15$,  $s$($\pi, \pi, \pi$) saturates to a finite value for $L\rightarrow\infty$, 
while it approaches zero for $x=0.15$. This analysis shows that indeed in 
the thermodynamic limit the AF order survives below the classical percolation 
threshold $x_p^{sc}$$\sim$0.31. Inset(ii) shows that the long range AF order
also persists up to $x$=0.15 for $T=0.01$.

In order to clarify the nature of the antiferromagnetic order below the 
$x_p^{sc}$$\sim$0.31, in Fig.~\ref{fig_mom}(a) we present the distribution 
of local moments in real space that includes both the $U=0$ and $U=8$ sites 
at $T=0.02$. The local moment distribution $P_{q}(M) = \sum_{M_i} \delta(M-M_i)$ 
show two peaks. In the limiting case of $x=1$ the expected plot is that of 
a single peak at $M=0.86$, implying uniform local moment magnitudes on all sites 
within the semi-classical calculation. As $x$ is reduced, a new peak at
lower $M$ appears that indicates the moment formation at $U=0$ sites. Inset of 
Fig.~\ref{fig_mom}(a) for $x=0.4$ confirms this scenario. This also corroborates 
the data shown in the inset of Fig.~\ref{fig_af} (a) 
that shows the signature of induced moment on the uncorrelated ($U=0$) sites.
Just as seen in Fig.~\ref{fig_af} (a), here too we see that if the $x$ reduces,
the location of the new peak moves towards $M=0.5$ signaling that 
the induced moment magnitudes get weaker. In addition, we note that the 
increase in the peak height at low $M$ is simply due
to increasing number of $U=0$ sites as $x$ approaches zero. 

Correlating the low temperature induced $M$ in the $U=0$ sites with the AF order
parameter in Fig.~\ref{fig_mom}(b), we see a crucial fact that the long range AF
order between $U=8$ sites depends on the existence of local moments at $U=0$ sites. 
Induced moment at $U=0$ sites are almost zero up to $x=0.15$ and the system remains
paramagnetic. Beyond $x=0.15$ the induced moment at $U=0$ sites increase and as 
a result the system enters into an antiferromagnetic state at small $x$. This
intimate relation between long range AF order and induced moments in the $U=0$
sites is central to the stability of the AF order below the classical percolation
threshold. Above $x \sim 0.31$, while the cooperation continues to exist as
discussed in context of Fig.~\ref{fig_af} (d), the AF order can stabilize by
the usual percolation mechanism as well.

We now analyze the recovery of the classical percolation threshold in two cases, 
(i) at relatively large temperature and (ii) at high U values.
In Fig.~\ref{fig_mom}(c) we show the AF order parameter as function of $x$ at three 
different temperatures. While there is a overall suppression of the order parameter 
magnitude with temperature, importantly we see that the critical threshold of $x$
for sustaining AF order moves towards the $x_p^{sc}$. The increase in temperature,
disrupts the local moment order for the $U=0$ sites, as was also seen from the
smaller charge gap in the DOS shown in  Fig.~\ref{fig_mit} (c). Thus the system
approaches the classical threshold for maintaining long range AF order. The percolation 
threshold $x_p$ also increases with $U$ as shown in Fig~\ref{fig_mom}(d). It is
known that at large $U$ the charge fluctuations are suppressed and the Hubbard model
is well described by the Heisenberg Model. The induced magnetic moment that
mediates the antiferromagnetism below $x_{p}^{sc}$ values (for example
induced moments at U=0 sites for $x$=0.2 and $U=8$ case) are not induced at
larger $U$ values [see the inset of Fig~\ref{fig_mom}(d)] due to suppression of
spin fluctuations. As a result the $x_{p}$ increases with $U$. 

\section{Conclusions}

We have employed a semi-classical technique to map out the temperature vs 
dilution phase diagram of the `diluted Hubbard model'. Our results at low 
temperature is close to unrestricted Hartree-Fock method and become progressively 
accurate with temperature. Within this scheme we have shown that away 
from the weak coupling ($U<<BW$) and strong coupling  ($U>>BW$) limit, site 
dilution weakens the long range magnetic order, but allows it to survive to 
dilution values much below the classical percolation threshold. At low temperature, 
the Mott insulator at $x=1$ evolves in to pseudo-gapped metal (for $x\leq 0.15$) 
by non-trivial spectral weight transfer phenomena that transforms the two Mott 
sub-bands into four sub-bands. Our analysis shows that in this regime the system 
has two distinct energy scales for charge excitations, one controlled by $U$ 
and another emergent gap that arises out of weak local moment induced on 
the $U=0$ sites. Such behavior of DOS qualitatively 
agrees with DQMC studies in two dimensions~\cite{ulmke}. By performing finite 
size scaling analysis, we also show that the induced moments at the $U=0$ 
sites and the $U\neq 0$ sites cooperate to form long range magnetic order in 
the thermodynamic limit. We further demonstrate that cooperation between the 
$U=0$ and $U\neq 0$ sites is crucial to the magnetic order by showing that 
increasing $U$ to large values, brings the critical percolation threshold 
to the classical value which requires system spanning AF patches exclusively 
made out of $U\neq 0$ sites. In addition by increasing temperature we have 
shown that the closure of the Mott gap by closing the smaller charge gap 
again disrupts the cooperation between the AF order between the two kinds 
of sites which pushes the percolation threshold to the classical values. 
This phenomenology is also seen in exact diagonalization where local Kondo 
coupling and RKKY scales compete to control the magnetic properties of $s-d$ 
models for carbon nanotubes \cite{luo} and broadly agrees with DQMC study 
on Lieb lattice \cite{lima}.

\section{Supplementary Information}

We consider the following electron-hole symmetric (EHS) one band Hubbard
Hamiltonian:

\begin{eqnarray}
H= -t\sum_{<i,j>,\sigma} c^{\dagger}_{i,\sigma}c_{j,\sigma} + U\sum_{i} \Big(n_{i,\uparrow}-\frac{1}{2}\Big) \Big(n_{i,\downarrow}-\frac{1}{2}\Big) \nonumber
\end{eqnarray}
\noindent

where `t' is the nearest neighbor hopping parameter and `U' denotes the
on-site Hubbard interaction. We set t=1 in our calculations.

After some trivial algebra and dropping a constant
term the EHS Hubbard model becomes:  

\begin{eqnarray}
H = -t\sum_{<i,j>,\sigma} c^{\dagger}_{i,\sigma}c_{j,\sigma} + U\sum_{i} n_{i,\uparrow}n_{i,\downarrow} -\frac{U}{2} \sum_{i} n_{i} \nonumber
\end{eqnarray}

We denote the nearest neighbor hopping term and the third term
which is a one body operator as $H_{0}$ and the second term which
is the interaction term as $H_{1}$.
We need to transform the interaction term as a combination of two quadratic
terms to set up the Hubbard-Stratonovich (HS) decomposition formalism.

\begin{eqnarray}
n_{i,\uparrow}n_{i,\downarrow} = \Big[ \frac{1}{4} n_{i}^{2} - S_{iz}^{2} \Big] = \Big[ \frac{1}{4} n_{i}^{2} - (\textbf{S}_{i}.\hat{\Omega}_{i})^{2} \Big]	
\end{eqnarray}

Here $\textbf{S}_{i}$ is the spin operator which is defined
as $\textbf{S}_{i}=\frac{\hbar}{2} \sum_{\alpha \beta}
c_{i,\alpha}^{\dagger} \sigma_{\alpha,\beta}
c_{i,\beta} $ , $\hat{\Omega}$ is an arbitrary unit vector,
$\sigma's$ are the Pauli matrices and we take $\hbar$
to be 1. We use the rotational invariance of $S_{iz}^{2}$  i.e.
$(\textbf{S}_{i}.\hat{\Omega}_{i})^{2} = S_{ix}^{2} = S_{iy}^{2} =
S_{iz}^{2}$ .  \\

Partition function for the Hamiltonian is $Z = Tr e^{-\beta H}$, where
$\beta$ is inverse temperature $(1/T)$ [$k_{B}$ is set to 1]. Next
we divide the interval $[0, \beta]$ into $M$ equally spaced slices, 
defined by $\beta = M \Delta \tau$, separated by $\Delta \tau$ and
labeled from 1 to M . For large $M$, $\Delta \tau$ is a small parameter
and allows us to employ the Suzuki-Trotter decomposition, so that we
can write $e^{-\beta(H_{0} +H_{1} )} = (e^{- \Delta \tau H_{0}}
e^{- \Delta \tau H_{1}} )^{M}$ to first order in $\Delta \tau$ .
Then using Hubbard-Stratonovich identity $e^{-\Delta \tau U [\sum_{i}
\frac{1}{4} n_{i}^{2} - (\textbf{S}_{i}.\hat{\Omega}_{i})^{2}]}$ can
be shown to be proportional to,\\

$\int d\phi_{i}(l) d\Delta_{i}(l) d^{2}\Omega_{i}(l) \times \\
\hspace*{3cm} e^{- \Delta \tau [\sum_{i}(\frac{\phi_{i}^{2}}{U}
+ i \phi_{i}(l)n_{i} + \frac{\Delta_{i}^{2}}{U} - 
2\Delta_{i}(l)\hat{\Omega}_{i}(l) \textbf{S}_{i}  )]}$ \\

Here $\phi_{i}(l)$ is the auxiliary field for charge density and
$\Delta_{i}(l)$ is auxiliary field for spin density and `(l)' is
a generic time slice. Further we define a new vector auxiliary
field $\textbf{m}_{i}$ as the product of $\Delta_{i}(l) \hat{\Omega}_{i}(l)$ .
Putting all of this back to the partition function we find the
effective Hamiltonian. Now we make two approximations which make
our model different from DQMC. Firstly we drop the $\tau$ dependence
of the Hamiltonian and we use the saddle point value of
$i\phi_{i} = \frac{U}{2} \textless n_{i} \textgreater $ . Lastly
by re-scaling $\textbf{m}_{i} \rightarrow (U/2)\textbf{m}_{i}$
we find the effective Hamiltonian as:

\begin{eqnarray}
H_{eff}= -t\sum_{<i,j>,\sigma} c^{\dagger}_{i,\sigma}c_{j,\sigma} 
+ U/2 \sum_{i} (\textless n_i \textgreater n_i - \textbf{m}_i . \sigma_{i})  \nonumber \\
  + (U/4)\sum_{i}(\textbf{m}_{i}^2 - {\textless n_i \textgreater}^2) -\frac{U}{2} \sum_{i} n_{i}  -\mu \sum_{i} n_i \nonumber
\end{eqnarray}

The chemical potential $\mu$ is used to tune the global electron
density equal to 1. In our calculation we have considered finite $U$
at randomly chosen sites $k$ with a concentration $x$ and $U=0$ at rest
of the sites (concentration $1-x$). So, we consider following diluted
Hamiltonian for our calculations:

\begin{eqnarray}
H_{eff}= -t\sum_{<i,j>,\sigma} c^{\dagger}_{i,\sigma}c_{j,\sigma} 
+ U/2 \sum_{k} (\textless n_k \textgreater n_k - \textbf{m}_k . \sigma_{k})  \nonumber \\
  + (U/4)\sum_{k}(\textbf{m}_{k}^2 - {\textless n_k \textgreater}^2) -\frac{U}{2} \sum_{k} n_{k}  -\mu \sum_{i} n_i \nonumber
\end{eqnarray}

\noindent \\
Acknowledgment: We acknowledge use of Meghnad2019 computer cluster at SINP.

\end{document}